# Monolithic Elastic Metasurface Design for Advanced Wave Manipulation via a Direct Wave-Shaping Topology Optimization Approach

Chun Min Li,[1] Wenjing Ye[1*]

[1]Department of Mechanical and Aerospace Engineering, The Hong Kong University of Science and Technology

Elastic metasurfaces offer precise control over elastic waves for applications such as vibration isolation, sensing, and imaging. However, achieving high-efficiency and scattering-free performance with complex functionalities remains a fundamental challenge. While conventional Generalized Snell's Law (GSL) designs suffer from inherent inefficiencies and parasitic scattering, recent alternatives—including impedance-matching methods and diffraction-grating-based metagratings—have sought to overcome these drawbacks. While efficiency improvement has been demonstrated, both methods are limited in generating complex wavefields such as focusing. Here, we propose a direct wave-shaping (DWS) topology optimization framework that bypasses these intermediate concepts and automates the design of high-performance, monolithic elastic metasurfaces. To mitigate the high computational expense, we parameterize the geometry with movable, deformable and interactable elliptical voids, thereby drastically reducing the design variables while enabling holistic optimization that inherently accounts for nonlocal inter-cell couplings. We demonstrate the framework by designing metasurfaces for challenging tasks including high-efficiency longitudinal-to-transverse wave conversion with large-angle beam steering, wavelength-multiplexed beam steering, and both reflective and transmissive metalenses with numerical apertures exceeding 0.99. Compared to state-of-the-art gradient-index, impedance-based, and hybrid designs, our metasurfaces consistently exhibit superior efficiency, significantly reduced spurious scattering, and enhanced focusing capability, with the most demanding transmissive metalens validated experimentally. This work establishes a scalable and computationally efficient pathway to realizing practical, high-performance elastic metasurfaces for advanced wave manipulation.

## I. INTRODUCTION.

Elastic metasurfaces are artificially engineered subwavelength structures capable of manipulating elastic waves in unprecedented ways. By carefully designing the geometry and material distribution of their subwavelength unit cells, these devices can achieve tailored wavefront shaping, including beam steering [1-4], focusing [2-4], wave entrapment [5], and mode conversion [6, 7]. These capabilities have enabled a wide range of applications, from vibration isolation and structural health monitoring to ultrasonic imaging. The performance of these applications, however, is critically dependent on the metasurface's efficiency and its ability to generate a pure, scattering-free wavefield, metrics on which conventional design methodologies often fall short.

The traditional design paradigm for elastic metasurfaces largely relies on Generalized Snell's Law (GSL) [8], which provides a phase modulation approach to control wave refraction and reflection [1-7]. While conceptually simple and widely adopted, it is well recognized that phase-gradient metasurfaces with local phase compensation have inherent limitations [9-10]. A fundamental issue is their typically low energy efficiency, especially under large-angle refraction conditions. This inefficiency arises from parasitic scattering into unwanted diffraction orders and the conversion of energy into surface waves. Furthermore, for metasurfaces fabricated from a single-phase material, the necessary perforations naturally weaken the structure's local density and stiffness, leading to a high impedance mismatch with the host material and consequently low energy transmission [11-13]. Even with additional design objectives to improve the transmission coefficient of individual unit cells, the resulting assembled metasurfaces often exhibit significant spurious waves, which drastically reduce the overall efficiency and purity of the output wavefield [1, 14]. The inability of GSL-based designs to consistently achieve high efficiency and scattering-free operation represents a major bottleneck for their practical application.

To overcome these challenges, alternative design approaches have emerged. Impedance-based design focuses on matching the effective impedance of the metasurface to a targeted impedance profile derived from the desired wave field in the host medium, thereby improving wave transmission/reflection efficiency. While this approach has been successfully applied in acoustic metasurfaces for high-efficiency wave manipulation [15-18], its extension to elastic systems is more challenging. A recent study sought to apply it to elastic metasurfaces for mode conversion and preservation of in-plane elastic waves [19]. Although the resulting designs exhibited very good performance compared to those designed using traditional GSL methodologies, spurious waves were observed, and full wavefront manipulation has not yet been fully achieved. These limitations are attributed to the complex interactions between multiple wave modes in elastic materials, which make it difficult to derive and implement the precise impedance profile required for high-efficiency control of complex wave fields.

*Contact author: mewye@ust.hk

Diffraction-grating-theory-based methods, which explicitly manipulate individual diffraction orders via periodic metagratings, have demonstrated highly efficient anomalous reflection and refraction for elastic waves [20, 21]. However, their capacity for creating complex wave functions is limited; for instance, they show poor performance for wave focusing [22]. This shortcoming has spurred the development of hybrid designs that combine GSL metasurfaces with metagratings, leading to significant performance improvement [22, 23].

A more direct and promising alternative is the direct wave-shaping (DWS) approach, which designs metasurfaces by setting the desired wave fields as the direct optimization objective, bypassing the need for intermediary concepts like phase gradients or specific impedance profiles [18]. Additionally, this method treats the metasurface as a monolithic structure rather than a periodic arrangement of independent unit cells. As such, this formulation naturally accounts for inter-cell couplings, thereby unlocking a broader design space and greater flexibility for achieving complex functionalities. It has yielded high-performance acoustic metasurfaces comparable to impedance-based designs [18]. Crucially, unlike impedance-based methods, this non-local approach extends straightforwardly to elastic waves. The primary obstacle, however, has been the prohibitive computational cost. Current implementations use density-based topology optimization (TO) [18], which requires a vast number of design variables to achieve high performance, leading to prolonged optimization times. These demands are exacerbated for elastic waves due to their vectorial nature and mode-converting interactions, making high-fidelity, monolithic design computationally intractable with conventional TO.

In this work, we present an efficient TO-based framework for the direct wave-shaping design of high-performance elastic metasurfaces that directly addresses the needs for high efficiency and scattering-free operation. Inspired by the Moving Morphable Component (MMC) framework [24], we parameterize the metasurface geometry using a set of movable and deformable elliptical hollow cylinders. This explicit geometric parameterization drastically reduces the number of design variables compared to conventional density-based TO, making the monolithic optimization of elastic metasurfaces computationally feasible while retaining ample structural flexibility. Leveraging these geometric primitives, we develop a multi-objective inverse design framework that directly engineers the transmitted and/or reflected wave fields to achieve target functionalities with high fidelity.

The effectiveness of our approach is demonstrated through several challenging test cases that highlight its superior performance:
• High-efficiency longitudinal-to-transverse wave conversion with large-angle beam steering.
• Wavelength-multiplexed wave steering with distinct refraction angles.
• High-efficiency, scattering-free reflective focusing with a large numerical aperture (NA > 0.9).
• High-efficiency, scattering-free transmission-type focusing with a large numerical aperture (NA > 0.9).

In all cases, our designs demonstrate a significant reduction in parasitic scattering and a marked improvement in energy efficiency compared to state-of-the-art methods, with the most challenging case (transmission focusing) being experimentally validated. This work establishes a robust and efficient pathway to realizing high-performance elastic metasurfaces that meet the stringent requirements of real-world applications.

## II. METHODS

In this study, we address the problem of elastic wave manipulation. Specifically, we consider a scenario where a longitudinal or transverse plane wave incident on a metasurface at an angle, $\theta_i$. The metasurface then performs the desired wave manipulation. An example is illustrated in Fig. 1, where part of an incident wave is refracted as a plane wave propagating at an angle $\theta_t$, while another part is reflected at an angle $\theta_r$. A coordinate system is defined with the x-axis aligned parallel to the metasurface. In the following subsections, we briefly introduce GSL and diffraction grating theory, followed by the detailed description of our proposed direct wave shaping (DWS) design approach.

*Contact author: mewye@ust.hk

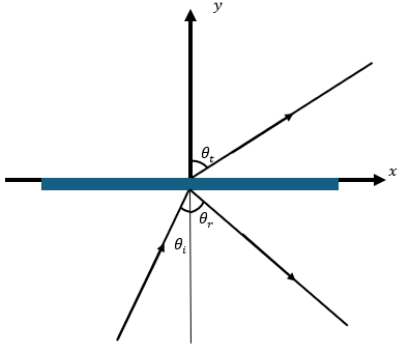

FIG. 1. Schematic of elastic wave manipulation with beam steering as an example.

### A. The Generalized Snell's Law

The GSL describes the refracted/reflected angle of a wave as it transmits/reflects through an interface that introduces an abrupt phase shift profile. It states that when a plane wave, characterized by a wavenumber $k_i$, is incident at an angle $\theta_i$ relative to the interface, the propagation angles $\theta_t$ and $\theta_r$ of the refracted wave and reflected wave are related to the phase shift $\phi(x)$ by the following expression:

$$k_r \sin \theta_r = k_t \sin \theta_t = k_i \sin \theta_i + \frac{d\phi}{dx}, \quad (1)$$

where $k_t$ and $k_r$ are the wavenumbers of the refracted wave and the reflected wave respectively. It is assumed that the phase profile $\phi(x)$ is smooth. If this assumption holds, it has been demonstrated that all transmitted waves adhere to the GSL, and no parasitic waves are generated [10]. However, if the phase profile lacks smoothness, parasitic components corresponding to other diffraction orders may arise.

Moreover, this law does not guarantee that the entire wave power will be transmitted through the interface. In particular, when the phase gradient is large, power losses may occur as some waves transform into surface waves. Additionally, in the context of elastic waves, mode conversion occurs as the wave propagates through the interface. The GSL only describes the direction of the wave transmitted through the interface and cannot be used to predict the amount of wave transmission and mode conversion through the interface.

### B. Diffraction Grating Equation

When a wave impinges on a periodic structure, it is diffracted into multiple directions as shown in Fig. 2. The propagation directions of these waves are described by the diffraction grating equation (Eq. (2)):

$$k_r \sin \theta_r^{(m)} = k_t \sin \theta_t^{(m)}$$
$$= k_i \sin \theta_i + m \frac{2\pi}{d}, \quad (2)$$

where $d$ is the period of the structure, $m$ is the diffraction order of the wave, and $\theta_t^{(m)}$ is the corresponding propagation direction.

The GSL can be viewed as a special case of the diffraction grating equation [25]. Specifically, when a linear, periodic phase profile that ranges from 0 to $2\pi$ is considered, the GSL reduces to Eq. (2) with $m = 1$. In other words, the GSL exclusively describes the first-order diffracted waves, but diffracted components of other integer orders also coexist in the wavefield.

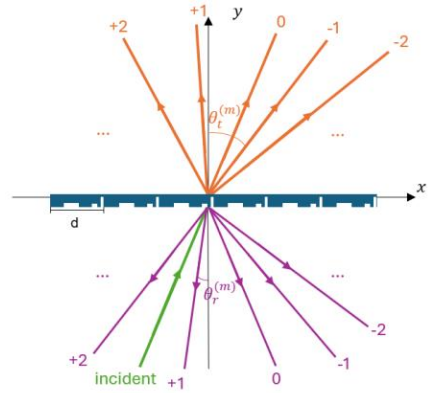

FIG. 2. Schematic of a wave diffracted by a periodic structure.

### C. DWS Design Framework

*1. Design representation and problem formulation*

In our DWS design framework, the 2D design domain is parametrized using $N$ elliptical holes embedded inside a rectangular domain as depicted in Fig. 3. Each ellipse is fully described by its semi-major axis $a_i$, semi-minor axis $b_i$, the angle of rotation $t_i$, and the coordinates of the center ($x_i$, $y_i$). These five parameters form the design variables for that ellipse. Consequently, the total number of design variables is $5N$. During the optimization process, the ellipses are free to move and interact, enabling the formation of large voids with more complex shapes as shown in Fig. 3. For 3D designs, ellipsoids can be employed.

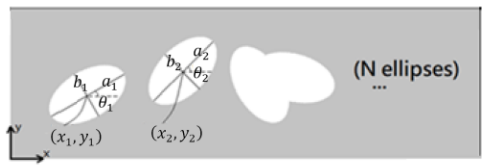

FIG. 3. Schematic of geometry parameterization used in the DWS framework.

*Contact author: mewye@ust.hk

The ranges for the design variables are defined as follows. Let $W$ and $H$ represent the width and height of the design domain, respectively, and $\lambda_L$ denote the wavelength of the longitudinal wave. The bounds for each design variable are assigned as: $a_i \in [0.01\lambda_L, H]$, $b_i \in [0.01\lambda_L, H]$, $\theta_i \in [0, 90°]$, $x_i \in [0, W/N_s]$, and $y_i \in [0, H]$. The lower bounds for $a_i$ and $b_i$ ensure that the holes are not too small for manufacturing. To reduce the design space, the movement of each ellipse is constrained with a dedicated subdomain with width of $W/N_s$, where $N_s$ is the number of subdomains; this directly defines the upper bound for $x_i$. All variables are normalized to the range of [0, 1], indicated by a tilde. For example, $a_i$ is normalized as $\tilde{a}_i = (a_i - 0.01\lambda_L)/(H - 0.01\lambda_L)$. The full set of optimization variables is then represented as $\boldsymbol{p} = (\widetilde{a_1}, \widetilde{b_1}, \widetilde{\theta_1}, \widetilde{x_1}, \widetilde{y_1}, \ldots, \widetilde{a_N}, \widetilde{b_N}, \widetilde{\theta_N}, \widetilde{x_N}, \widetilde{y_N}) \in [0,1]^{5N}$.

The design problem is formulated as an optimization problem in Eq. (3), with desired wave control functions defined as the design objectives. In this DWS approach, the objective function is constructed directly from the displacement wavefield, computed via the finite element simulations outlined in Section II.C.3. For a metasurface design problem involving K performance measures, the optimization is expressed as follows:

$$\min \boldsymbol{J}(\vec{u}(\boldsymbol{p})) = \begin{bmatrix} J_1(\vec{u}(\boldsymbol{p})) \\ J_2(\vec{u}(\boldsymbol{p})) \\ \vdots \\ J_K(\vec{u}(\boldsymbol{p})) \end{bmatrix}, \quad (3)$$

subjected to

$$\rho \frac{\partial^2 \vec{u}}{\partial t^2} = \nabla \cdot \sigma + \vec{F}, \quad (4)$$
$$\text{ISP}(\boldsymbol{p}) = 0, \quad (5)$$
$$\text{TC}(\boldsymbol{p}) = 0. \quad (6)$$

Here, Eq. (4) is the governing equation of linear elasticity, where $\rho$ is the density, $t$ is the time, $\sigma$ is the stress tensor field, and $\vec{F}$ is the body force field. Eq. (5) and Eq. (6) describe the geometrical constraints. In particular, ISP denotes the number of isolated solid parts and TC denotes the number of thin structures with width less than 0.2mm. These thin structures form when two elliptical holes are close but not intersecting. Both features are undesirable because elastic waves cannot propagate into isolated solid parts, and thin connections cause inaccurate simulation results and pose manufacturing challenges.

*2. Optimization method*

The optimization problem is solved using Adaptive Geometry Estimation-based Multi-objective Evolutionary Algorithm (AGE-MOEA) [26]. AGE-MOEA is a novel evolutionary algorithm that utilizes an adaptive diversity metric accounting for the geometry of the estimated Pareto front. Unlike traditional evolutionary algorithms like Non-dominated Sorting Genetic Algorithm II (NSGA-II), which rely on a fixed Manhattan distance and implicitly assume a linear and symmetric Pareto-front geometries, AGE-MOEA dynamically adjusts its diversity metric to better represent real-world problems. This innovation allows AGE-MOEA to excel at finding high-performing and diverse sets of solutions, thus enabling consistent performance across diverse optimization problems, including those with high-dimensional objective spaces.

AGE-MOEA follows the general framework of the canonical NSGA-II. At each generation $n$, a parent population $P_n$ of size $N$ is used to create an offspring population $Q_n$ of size $N$ through standard genetic operators such as simulated binary crossover (SBX) and polynomial mutation. The combined population $R_n = P_n \cup Q_n$ is then sorted into a hierarchy of non-dominated fronts ($F_1, F_2, F_3, \ldots$) using the fast non-dominated sorting algorithm. The first front $F_1$ represents the best available approximation of the Pareto front for the current generation.

To analyze the geometry of the Pareto front, AGE-MOEA normalizes the objective values of all solutions within $F_1$. This normalization uses the minimum and maximum values of each objective in $F_1$, effectively projecting the front into a unit hypercube where the ideal point is at the origin (0, 0, ..., 0) and the nadir point is near the corner (1, 1, ..., 1).

The geometry of the Pareto front is then estimated by identifying the point in the normalized front $F_1$ that is farthest from the ideal point, measured using the Euclidean distance. This point is treated as the corner of the front's hypervolume. The algorithm then selects the value of p that minimizes the Minkowski p-distance between this extreme point and its nearest neighbor within $F_1$. This adaptively chosen p-value reflects the geometry of the Pareto front, indicating whether it is linear (p = 1), spherical (p = 2), or concave (a high p). The adaptively chosen p-norm is subsequently used to calculate a novel survival score for each solution in $R_t$. This score balances convergence and diversity. The convergence component is quantified as the p-norm distance between a solution's normalized objective vector and the ideal point, while the diversity component is calculated as the p-norm distance to the solution's nearest neighbor within the same non-dominated front. These components are combined into a single survival score, which is used for environmental selection. After filling the population size limit with the best-performing non-dominated fronts, any remaining slots are allocated to individuals with the lowest survival scores from the next front. This ensures that both convergence and diversity are consistently promoted throughout the evolutionary process, even for complex Pareto-front geometries.

*Contact author: mewye@ust.hk

In this study, AGE-MOEA is implemented using the Pymoo package [27]. The population size is set to 50, and 50 offspring are generated per generation. The SBX scheme with η = 10 and a crossover probability of 0.95 is used, along with polynomial mutation (η = 10). The optimization is performed iteratively until the Pareto surface stabilizes, defined as no improvement for over 200 generations.

### 3. Performance and constrain evaluation

The metasurface's performance is evaluated during optimization using finite element (FE) simulations to compute the displacement field. This field is post-processed to extract key metrics, including the longitudinal and transverse wavefields, power flux, and intensity. These procedures are detailed in this subsection.

The general simulation setup is depicted in Fig. 4. A line source is placed below the design domain of the metasurface to excite the elastic wave. To suppress reflections at the artificial boundaries constructed in FEA, an absorption layer with increasing damping (ALID) of thickness $h_1$ is applied. The ALID is implemented using Rayleigh damping, with the damping ratio defined as $\zeta(w) = \zeta_{\max}(h_1 - w)^2/h_1^2$, where $\zeta_{\max} = 0.8$ and $w$ is the distance from the boundary.

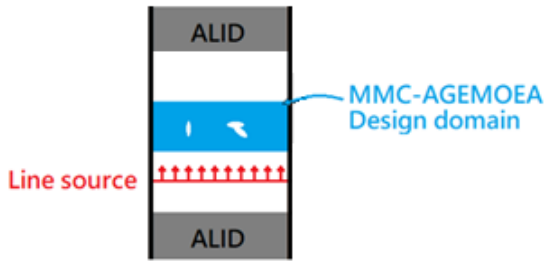

FIG. 4. Schematic of the general FE simulation setup.

The finite element simulations are performed in the frequency domain using COMSOL Multiphysics 6.3. The mesh size is set as $\lambda_T/5$ outside the design domain and $\lambda_T/20$ inside it. These values correspond approximately to $\lambda_L/9$ and $\lambda_L/35$ respectively if plane stress assumption is used.

The longitudinal and transverse wave components are extracted from the simulated displacement field by taking the divergence and curl of the displacement field as shown below. Without loss of generality, the displacement wavefield of a propagating plane wave can be expressed as the superposition of longitudinal and transverse waves as:

$$\vec{u} = \vec{u_L} + \vec{u_T} = A_L \vec{d_L} e^{j(\vec{k_L} \cdot \vec{x} - \omega t)} + A_T \vec{d_T} e^{j(\vec{k_T} \cdot \vec{x} - \omega t)}, \quad (7)$$

where $\vec{x}$ is the position vector, $A_L$ and $A_T$ are the amplitudes of the longitudinal wave and transverse wave, respectively. The wave vectors for the two wave types are given as $\vec{k_L} = k_L[\sin\theta_L, \cos\theta_L]^T$, $\vec{k_T} = k_T[\sin\theta_T, \cos\theta_T]^T$, and the respective particle displacement directions are $\vec{d_L} = [\sin\theta_L, \cos\theta_L]^T$, $\vec{d_T} = [\cos\theta_T, -\sin\theta_T]^T$. From this, it follows that

$$|\nabla \cdot \vec{u}|/k_L = A_L, \quad (8)$$

and

$$|\nabla \times \vec{u}|/k_T = A_T. \quad (9)$$

Efficiency is a critical performance measure for metasurfaces, and its definition is related to the power flux and intensity. The power flux of an elastic wave through a surface is given by $P = \sigma \vec{v}$ where $\sigma$ is the stress field, $\vec{v} = \partial \vec{u}/\partial t$ is the velocity field, and $\vec{n}$ is the surface normal vector. For a plane wave travelling across a planar surface, the power flux simplifies to

$$P = \frac{1}{2}\rho c \omega^2 A^2 \cos\theta, \quad (10)$$

where $\rho$ is the density, $c$ is the density, and $\omega$ is the angular frequency of the wave. Following the convention in the elastic wave community [28-32], intensity in this work is defined as the amplitude squared. For example, the intensity of the longitudinal wave is $I_L := A_L^2 = |\nabla \cdot \vec{u}|^2/k_L^2$ and it is not equivalent to the power flux.

To eliminate the isolated solid parts and thin connections discussed at the end of Section II.C.1, the quantities ISP and TC are counted explicitly. The design is first represented as a binary image, then it is passed to the ndimage.label function from the Scipy library, which can identify the number of isolated parts and number each of them, from that the ISP can be found. To obtain TC, the image is dilated with a kernel of size 0.2mm to expand hole regions. If two holes becomes one after dilation, it indicates that there is a thin connection between the two original holes. The counted ISP and TC are then fed to the Pymoo package for the geometrical constraint implementation.

### D. Material fabrication and experimental measurement

*1. Experimental setup and instrumentation*

*Contact author: mewye@ust.hk

To further validate the performance of the developed design framework, a representative metasurface design is fabricated and its performance is experimentally measured. The metasurface sample is fabricated on a 1mm-thick aluminium plate using a high-precision fibre laser cutting system (Glorystar GS-0605P). To suppress boundary reflections, the plate perimeter is treated with Blu-Tack. Elastic waves are generated using an array of 50 mm × 50 mm PZT-5H transducers. Surface velocity is measured with a Polytec OFV-505 laser Doppler vibrometer (LDV), and retroreflective tape is applied at all measurement points to improve optical return. The PZT array is driven by a Keysight 33500B function generator, and no power amplifier is used. The LDV output is recorded with a Teledyne LeCroy HDO6054A oscilloscope. Photographs of the specimen and LDV are provided in Fig. 5.

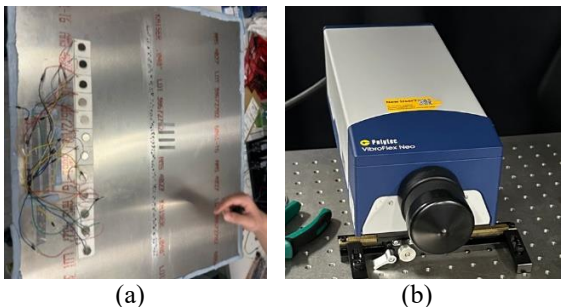

(a)            (b)

FIG. 5. (a) Metasurface specimen and (b) Polytec OFV-505 LDV.

*2. Wave-mode selection and LDV configuration*

The target response is the fundamental symmetric Lamb mode (S0). Because the antisymmetric Lamb mode (A0) and the shear-horizontal mode (SH) may also be present, the measurement geometry and timing are configured to isolate S0. The plate is mounted vertically (normal along the altitude direction) and oriented at 30° relative to the LDV beam in the azimuthal direction. In this geometry, tilting in azimuth enables measurement of S0, whose particle motion is predominantly along its horizontal propagation direction [33]. Minimization of SH sensitivity is achieved by keeping the plate normal to the laser in the altitude direction, because SH particle motion is primarily along that direction. Minimization of A0 sensitivity is achieved by only considering the first wave packet of the signal, during which the A0 wave has not yet arrived at the measurement point.

*3. Excitation, acquisition, and signal processing*

The excitation is a 9.8V, 200 kHz, 5-cycle, Hanning-windowed toneburst applied through the PZT array. The LDV bandwidth and velocity range are set to 500 kHz and 10 mm/s, respectively. The oscilloscope sampling frequency and record length are set to be 100 MHz and 1 ms, respectively. The oscilloscope is run in auto-acquisition mode, and one waveform is saved after every 10,000 samples are collected. At each measurement location, the excitation–measurement sequence is repeated 10 times, and the results are ensemble-averaged to improve signal-to-noise ratio. The signals are denoised using a Butterworth bandpass filter with 100 kHz and 400 kHz cutoff frequencies. The TOF is defined as the first time instant at which the signal amplitude exceeds 1 mV, an empirically determined threshold that discriminates the physical response from ambient noise. After aligning the 15 μs time window to the TOF, the longitudinal-wave magnitude at each location is taken as the maximum absolute amplitude within the window.

## III. RESULTS

This section presents the design and performance evaluation of two benchmark metasurface applications namely beam steering and wave focusing. For beam steering, we demonstrate both single-frequency steering of a longitudinal wave into a transverse wave and dual-frequency steering of two longitudinal waves into two transverse waves along different directions. For wave focusing, we design both transmitted and reflected metalens. The results are compared against existing methods where available.

### A. Simultaneous Mode Conversion and Beam Steering

The effectiveness of the DWS framework is demonstrated in this subsection through the design of 2D elastic metasurfaces that simultaneously convert normally incident longitudinal waves into transverse waves and steer their direction. The designed metasurfaces feature a periodic structure with periodic boundary conditions (PBC) applied on both lateral edges of the simulation domain, as shown in Fig. 6. To avoid the elliptical holes cut through the PBC and causes modelling complications, a thin region of thickness $t = 1$mm next to each PBC is defined as non-design region, and the elliptical holes that cross the region are trimmed. The period of the metasurface, $d$, is determined by Eq. (2) and depends on the targeted wave mode and transmission direction. ALID are applied to the top and bottom boundaries of the simulation domain.

*Contact author: mewye@ust.hk

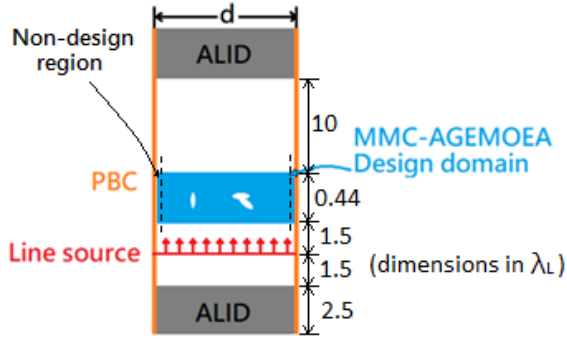

FIG. 6. Schematic of FE simulation setup for beam steering metasurface design evaluation.

*1. Single-Frequency Design*

For this application, the frequency of interest is 100kHz, and the model uses aluminum (E = 70GPa, $\rho$ = 2700kg/m$^3$ and $\nu$ = 0.33) under plane stress conditions. Two metasurfaces are designed to transmit waves at the targeted propagation angles $\theta_t = 33°$ and 60° respectively. Taking $m = 1$ in Eq. (2), the widths of the repeating unit $d$ are set to be 56.4mm and 36.0mm respectively. The number of elliptical holes in the metasurface is set to be 16.

For this design, only one objective function is defined:
$$J(\mathbf{p}) = -A_{T,\theta_t}/A_N, \quad (11)$$
where $A_{T,\theta_t}$ is the amplitude of the transmitted transverse wave component propagating at the desired direction $\theta_t$, and $A_N$ is a normalization factor defined as the theoretical limit of $A_{T,\theta_t}$
$$A_N = A_0\sqrt{\cos\theta_t \, c_L/c_T}, \quad (12)$$
where $A_0$ is the amplitude of the incident wave, and $c_L$ and $c_T$ are the velocities of longitudinal and transverse wave respectively.

The optimized geometries of two designs together with their wave controlling performances are illustrated in Fig. 7. Visually, the results show that the incident longitudinal wave has been successfully converted into a transverse wave propagating in the desired direction, with minimal presence of other diffraction orders. To quantify the performance, the normalized wave powers of the transmitted and reflected wavefields for various orders are calculated using Eq. (10) and are summarized in Table 1. The efficiency of the metasurfaces designed for 33° and 60° transmission angles is 91.6% and 93.4%, respectively. Only minimal power is lost, primarily due to reflections or refractions of different orders.

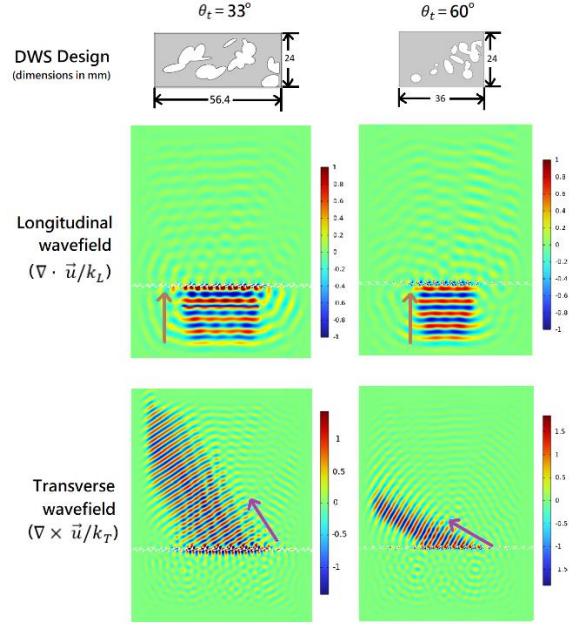

FIG. 7. DWS designs for targeted propagation angles $\theta_t$ = 33° and 60°, and the corresponding wavefields as a longitudinal wave incident normally from the bottom. The amplitudes are normalized by the amplitude of the incident wave.

For comparison, a metasurface is designed based on GSL as outlined in [14]. In this approach, the periodic structure (supercell) contains eight unit cells. Each unit cell is designed separately to convert a longitudinal wave into a transverse wave while imposing the phase shift designated by the GSL. These cells are then assembled to form the metasurface. Specifically, the dimensions of the unit cells are 24 mm × 7.05 mm, with a 1 mm gap between unit cells to avoid interaction between unit cells. This results in a design domain size of 24 mm × 6.05 mm for each unit cell. These unit cells are designed using the proposed TO method. Each unit cell contains three ellipses (N = 3), and the optimal parameters are determined by simultaneously maximizing the transmission coefficient and minimizing the difference between the phase shift and the targeted phase shift. Mathematically, the objective functions are defined as
$$J_1(\mathbf{p}) = -A_T/(A_0\sqrt{c_L/c_T}) \quad (13)$$
$$J_2^{(n)}(\mathbf{p}) = \frac{1}{\pi}\left|\phi - \frac{n\pi}{4}\right|,$$
where $A_T$ is the amplitude of the transmitted transverse wave, $A_0$ is the amplitude of the incident wave, $\phi$ is the phase shift across the unit cell, and $\frac{n\pi}{4}, n = 0,1,\cdots 7$, denotes the targeted phase shift for each cell. The corresponding wavefields resulting from a longitudinal wave incident on individual unit cell are assembled and plotted in Fig. 8.

*Contact author: mewye@ust.hk

The detailed optimization results are presented in Supplementary Materials. The results indicate that the fraction of power transmitted through each unit cell as a transverse wave exceeds 91.6%. Additionally, the values of the second objective function $J_2^{(n)}$, are less than 0.03 radians, demonstrating that each designed unit cell achieves both high-quality transmission and precise phase control.

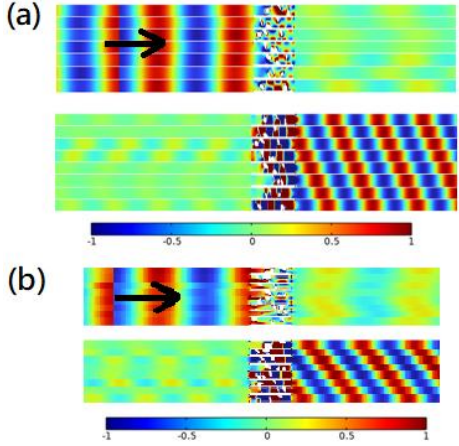

FIG. 8. Normalized longitudinal wavefield (top) and transverse wavefield (bottom) when a longitudinal wave incident on each unit cell designed for (a) $\theta_t = 33°$ and (b) $\theta_t = 60°$ from the left.

TABLE 1. Fraction of incident power that has been scattered by the beam steering and mode converting metasurfaces.

| Targeted $\theta_t$ | | 33° | | 60° | |
|---|---|---|---|---|---|
| Wave modes | m | DWS | GSL | DWS | GSL |
| Transmitted transverse | 1 | **91.6** | **69.0** | **93.4** | **13.4** |
| | -1 | 0.4 | 6.9 | 2.7 | 10.4 |
| | 0 | 0.3% | 1.5% | 0.8% | 4.1% |
| Transmitted longitudinal | 1 | 0.2% | 7.4% | - | - |
| | -1 | 0.2% | 2.9% | - | - |
| | 0 | 0.4% | 1.6% | 0.8% | 35.8% |
| Reflected longitudinal | 1 | 2.5% | 0.3% | - | - |
| | -1 | 0.1% | 1.4% | - | - |
| | 0 | 1.3% | 2.2% | 1.8% | 10.1% |
| Reflected transverse | 1 | 1.3% | 3.4% | 0.2% | 15.7% |
| | -1 | 0.1% | 2.2% | 0.3% | 4.0% |
| | 0 | 0.5% | 0.9% | 0.1% | 6.5% |

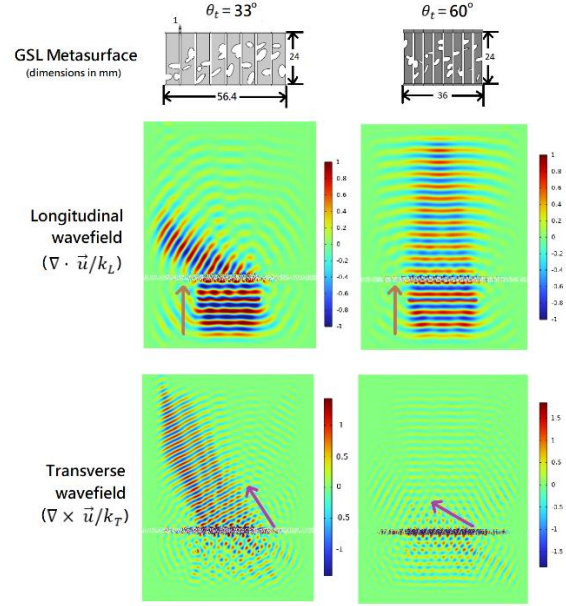

FIG. 9. GRIN metasurfaces for targeted propagation angles $\theta_t = 33°$ and 60°, and the corresponding wavefields as a longitudinal wave incident normally from the bottom.

However, when these unit cells are assembled to form a metasurface, its performance is less satisfactory. The wavefields, as a longitudinal wave is incident on the metasurface from the bottom, are shown in Fig. 9, and the fraction of incident power that is transmitted through or reflected by the metasurface is summarized in Table 1.

For the 33° metasurface, most of the incident wave is converted into a transverse wave propagating at 33°. However, the wavefronts are not straight, indicating the presence of other orders. Additionally, a portion of the wave transmits through the metasurface as a longitudinal wave, predominantly traveling at 72°, resulting in a relatively low total efficiency of 69.0%.

For the 60° metasurface, the efficiency is even lower, which achieves only 13.4%. Most of the wave fails to propagate in the desired direction and instead transforms into spurious waves. Notably, a significant portion, 35.8%, of the input wave power transmits through the metasurface as a longitudinal wave propagating in the normal direction. These results align with the known limitations of designs based on the GSL, which tend to exhibit inferior efficiency when the phase gradient is large. Similar findings have been reported in recent literature [19].

Despite satisfying the impedance matching condition individually—which ensures minimal longitudinal wave transmission through each unit cell—the specific impedance required to achieve unitary efficiency in beam steering differs. This difference arises because interactions between unit cells must be accounted for to minimize energy loss.

*Contact author: mewye@ust.hk

In contrast, our DWS framework addresses this issue by treating the metasurface as a single entity. The impedance profile required for the targeted wave manipulation task is determined by directly optimizing the associated physical objective functions, ensuring superior performance and efficiency.

### 2. Dual frequency design

A major drawback of passive metasurfaces is their lack of tunability. To address this, wavelength-multiplexed designs have emerged as a promising approach for achieving multiple functionalities. In this testing case, the metasurface is designed to convert a 50kHz normally incident longitudinal wave to a transverse wave travelling at $\theta_t^{(1)} = 30°$, and a 100kHz normally incident longitudinal wave to a transverse wave travelling at $\theta_t^{(2)} = -30°$. This model uses aluminum (E = 70.05GPa, $\rho$ = 2700kg/m³, $\nu$ = 0.33) under plane strain condition. The period of the metasurface is set as 124.8mm, so that waves traveling at $\theta_t^{(1)} = 30°$ and at $\theta_t^{(2)} = -30°$ correspond to the first-order and the negative second-order transmitted transverse waves, respectively. Since this metasurface is designed to have dual functions, the design complexity is higher than that of the first example shown in Fig. 7. Hence, the number of ellipses is set to be N = 24 to allow a larger design space.

The metasurface is optimized by maximizing the amplitude of the desired transmitted transverse wave components simultaneously for the two frequencies. The two objectives for this design problem are

$$J_1(\boldsymbol{p}) = -A^{(1)}_{T,\theta_t^{(1)}}/A_N^{(1)}, \tag{14}$$

and

$$J_2(\boldsymbol{p}) = -A^{(2)}_{T,\theta_t^{(2)}}/A_N^{(2)},$$

where $A^{(1)}_{T,\theta_t^{(1)}}$ and $A^{(2)}_{T,\theta_t^{(2)}}$ are the amplitudes of the desired wave components at the two frequencies, and $A_N^{(j)} = A_0\sqrt{\cos\theta_t^{(j)}\, c_L/c_T}$ are the normalization factors.

The proposed DWS design method produces a population of optimized designs. Each design in the population is optimal in the sense that no other design achieves higher power efficiencies in both frequencies simultaneously. Among these designs, the one with comparable efficiencies (78.1% and 73.3%) at the two frequencies are picked. While these values are lower than those of single-frequency DWS designs—reflecting the increased design complexity—they still exceed the performance of the single-frequency GSL design. The geometries of the design together with its wave controlling performances at the two frequencies are illustrated in Fig. 10, and the detailed design parameters are listed in Supplemental Materials.

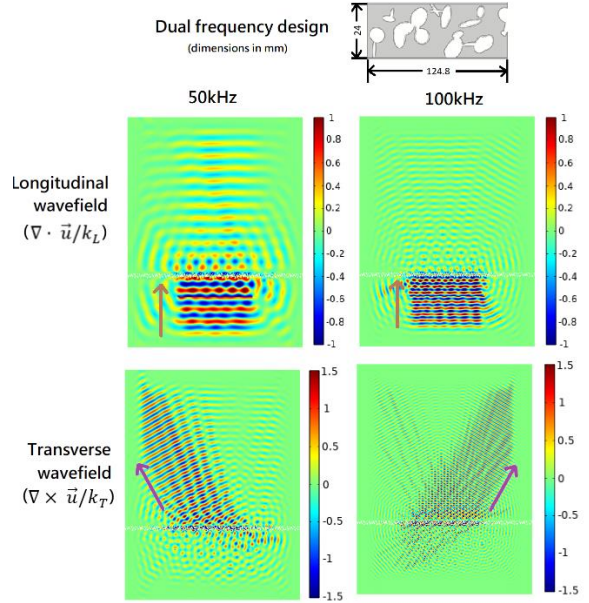

FIG. 10. DWS design for wavelength-multiplexed steering. A normally incident longitudinal wave from the bottom is refracted at distinct angles depending on the wavelength.

TABLE 2. Fraction of incident power scattered by the frequency-multiplexing, beam-steering and mode-converting metasurface shown in Fig. 10.

|  | 50kHz | 100kHz |
|---|---|---|
| Desired transmitted transverse component | **78.1%** | **73.3%** |
| Other transmitted transverse components | 3.0% | 8.0% |
| Transmitted longitudinal components | 8.4% | 4.3% |
| Reflected longitudinal components | 4.0% | 3.9% |
| Reflected transverse components | 6.3% | 10.4% |

### B. Metalenses

#### 1. Reflective Elastic Metalens

In this subsection, the effectiveness of the DWS design framework is demonstrated through two designs of reflective elastic metalenses with different numerical apertures. The metalenses, constructed from steel (E = 211GPa, $\rho = 7874$kg/m$^3$, $\nu = 0.29$), are modeled under plane strain condition. The width of the metasurfaces is $2W = 20.4\lambda_L$. The two metalenses are designed to focus a normally incident 200 kHz longitudinal wave at focal lengths of $4\lambda_L$ and $1.1\lambda_L$, corresponding to numerical apertures (NA) of 0.931 and 0.9942, respectively.

*Contact author: mewye@ust.hk

Due to the symmetric nature of the problem, the design is assumed to be symmetric about the centerline, making it sufficient to simulate only half of the metasurface to evaluate its performance. The FEM simulation setup for metasurface evaluation is shown in Fig. 11. A symmetry condition is imposed on the left edge, a fixed boundary condition is applied at the top edge. A thin region of thickness $t = 1$mm below the fixed top edge is a non-design region. ALID are applied on the right and bottom edges of the simulation domain.

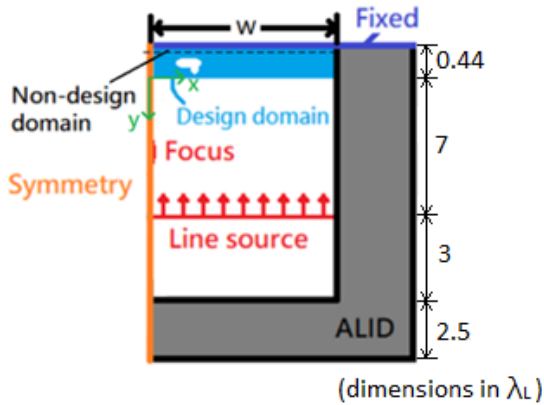

FIG. 11. Schematic of the FEM setup for the design and performance evaluation of reflective metalens

To evaluate the performance of reflective metasurfaces, it is necessary to obtain reflected wavefields from the simulated wavefields. For this purpose, the incident wavefield is determined using a baseline simulation in which the fixed boundary condition is replaced by an absorption layer and the metasurface is removed. The reflected longitudinal and transverse fields $B_L$ and $B_T$ are then calculated by subtracting the corresponding values in the baseline simulation from $\nabla \cdot \vec{u}/k_L$ and $\nabla \times \vec{u}/k_T$ respectively.

To design metalenses, the design domain is divided into 24 rectangular subdomains and each subdomain contains 8 ellipses, resulting in a total of N = 192 ellipses. Five objective functions, formulated based on the displacement field, are defined for this optimization problem, as shown in Eq. (15). $J_1$ is the full width at half maximum (FWHM), and it is minimized to improve the sharpness of the focusing. $J_2$ is the difference between the targeted focal length and the realized focal length by the metasurface. By minimizing $J_2$, the positional accuracy of the focusing can be enhanced. $J_3$ is the negative of the normalized power of longitudinal wave in the focusing region, which is a measure of the efficiency of the metasurface. $J_4$ is the ratio between the intensity in the focusing region and the intensity of the entire focal plane, which can be understood as a measure of signal to noise ratio (SNR). $J_5$ is the normalized intensity of transverse wavefield. By minimizing it, more power is channeled to longitudinal wave as desired. All objectives are normalized to ensure they are the same order of magnitude.

$$\begin{aligned}
J_1(\boldsymbol{p}) &= \text{FWHM}/\lambda_L, \\
J_2(\boldsymbol{p}) &= |F - F_0|/\lambda_L, \\
J_3(\boldsymbol{p}) &= -\frac{\int_0^{x_{\min}} |B_L(x, F_0)|^2 dx}{A_I^2 W}, \\
J_4(\boldsymbol{p}) &= -\frac{\int_0^{x_{\min}} |B_L(x, F_0)|^2 dx}{\int_0^\infty |B_L(x, F_0)|^2 dx}, \\
J_5(\boldsymbol{p}) &= \frac{\int_D |B_T(x, F_0)|^2 dx}{A_I^2 W},
\end{aligned} \quad (15)$$

where $F_0 = 4\lambda_L$ or $1.1\lambda_L$ is the desired focal length, $F$ is defined as the y-coordinate of the point where the longitudinal wavefield attains its maximum along its centerline:

$$F = \text{argmax}_y |B_L(0, y)|, \quad (16)$$

$A_I$ is the amplitude of the incident wave, $D$ is the whole domain omitting the metasurface, and $x_{\min}$ is the x-coordinate of the point on the focal plane where the amplitude achieves its first local minimum.

Optimizations are performed for the two focal lengths. For each case, two designs are selected from the resulting design population: (1) the design with the highest efficiency (lowest value for $J_3$), and (2) a design with high efficiency and signal-to-noise ratio (low $J_3$ and $J_4$). The geometries of the designs are shown in Fig. 12, and the corresponding parameters are listed in Supplementary Materials.

*Contact author: mewye@ust.hk

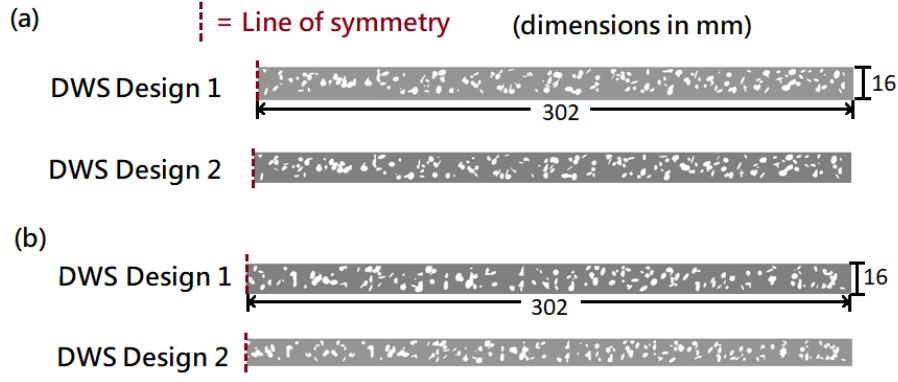

FIG. 12. DWS reflective metalenses designed to focus longitudinal wave at $F_0 = 4\lambda_L$ and $F_0 = 1.1\lambda_L$ respectively.

For comparison, GRadient-INdex (GRIN) metalens based on the GSL are designed for each target focal length. These metasurfaces have the same dimensions as the DWS designs, and their phase profile are given by

$$\phi(x) = 2\pi\left(\frac{-1}{\lambda}\left(\sqrt{x^2 + F_0^2} - F_0\right) \bmod 1 + 1\right), \quad (17)$$

where the phase shift profile for the case with $F_0 = 1.1\lambda_L$ is also depicted in Fig. 13.

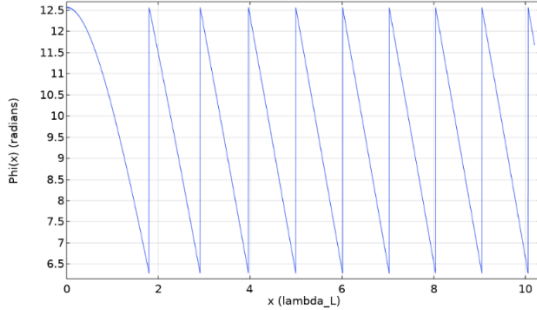

FIG. 13. Phase shift profile $\phi(x)$ obtained from GSL for the design case with $F_0 = 1.1\lambda_L$.

To realize the phase shift profile $\phi(x)$ while ensuring impedance matching with the host material, the density and Young's modulus of the material of metalenses are set as

$$\rho(x) = \frac{1}{2}\frac{\lambda_L}{H}\frac{\phi(x)}{2\pi}\rho_0, \quad (18)$$
$$E(x) = 2\frac{H}{\lambda_L}\frac{2\pi}{\phi(x)}E_0.$$

It should be noted that such functionally graded materials are unlikely to exist naturally. They are used in this study for a specific rationale: to determine the upper performance limit achievable by the GSL methodology. This provides a rigorous benchmark for evaluating the performance of our proposed DWS designs.

The focusing performances of our designs and the GRIN metalens are illustrated in Fig. 14 and Fig. 15. Moreover, the results along the focal plane ($y = F_0$) and the symmetry line ($x = 0$) are shown in Fig. 16. From the results plotted in Fig. 16, it can be seen that all designs focus the wave at the desired positions. For quantitative comparisons, performance metrics including the FWHM, efficiency, amplification factor, and SNR are summarized in Table 3 and 4, where efficiency is defined as

$$\eta = \frac{\int_{-3\text{FWHM}}^{3\text{FWHM}} |A_L(x,F)|^2 dx}{\int_{-\infty}^{\infty} |A_I(x,F)|^2 dx}, \quad (19)$$

following the literature, and IEF stands for intensity enhancement factor, which is the ratio of the intensity of the longitudinal field at the focal point to the incident intensity.

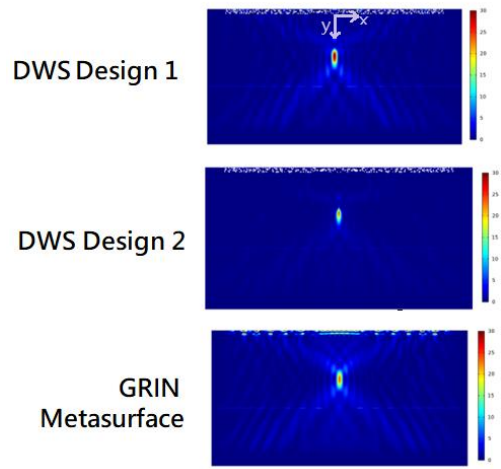

FIG. 14. The normalized intensity fields as a longitudinal wave incident normally from the bottom of the $F_0 = 4\lambda_L$ DWS designs and the comparison with a GRIN metasurface.

*Contact author: mewye@ust.hk

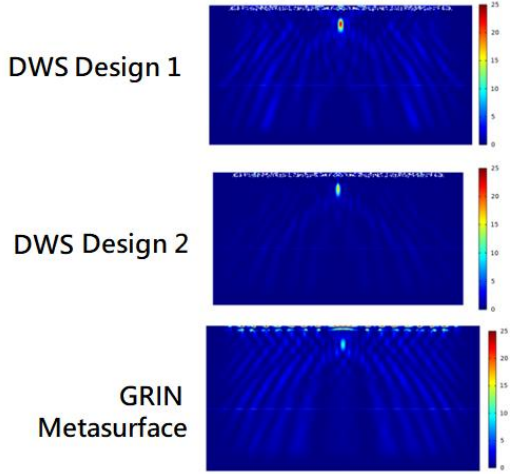

FIG. 15. The normalized intensity fields and the comparison with a GRIN metasurface for the design case $F_0 = 1.1\lambda_L$.

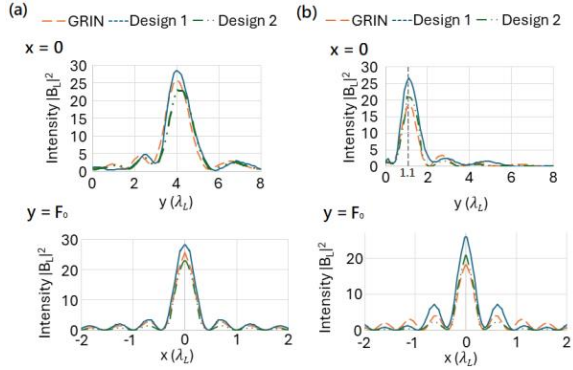

FIG. 16. Slices of the intensity fields along the line of symmetry x = 0 and and focal line y = F0 for the design case (a) $F_0 = 4\lambda_L$ and (b) $F_0 = 1.1\lambda_L$.

TABLE 3. Performance metrics for DWS and GRIN metalenses, and comparison with state-of-the-art hybrid metalens for the case of NA = 0.931

| Metalens | Design 1 | Design 2 | GRIN | J. Mei (2024) |
|---|---|---|---|---|
| $F_0(\lambda_L)$ | | 4 | | 20 |
| FWHM ($\lambda_L$) | 0.413 | 0.394 | 0.405 | 0.39 |
| Efficiency $\eta$ | **68.9%** | 55.8% | 63.6% | 54.6% |
| IEF per width ($1/\lambda_L$) | **1.383** | 1.110 | 1.250 | 1.103 |
| SNR(dB) | 9.29 | **12.38** | 9.63 | 8.13 |

TABLE 4. Performance metrics for DWS and GRIN metalenses, for the case of NA = 0.9942

| Metalens type | Design 1 | Design 2 | GRIN |
|---|---|---|---|
| FWHM($\lambda_L$) | 0.361 | 0.341 | 0.333 |
| Efficiency $\eta$ | **61.6%** | 47.6% | 42.6% |
| IEF | **26.0** | 20.1 | 18.0 |
| SNR(dB) | 5.58 | **9.90** | 6.50 |

*Contact author: mewye@ust.hk

For the case of lower NA (NA = 0.931, $F_0 = 4\lambda_L$), the focusing performances of DWS Design 1 and the GRIN metalens are similar. The DWS design achieves slightly higher efficiency and intensity enhancement factor (68.9% and 1.383) compared to the GRIN metalens (63.6% and 1.250), but it has a slightly larger FWHM (0.413 vs. 0.405) and a lower SNR (9.29 vs. 9.63). In other words, while Design 1 focuses more power near the focal spot, it exhibits a slight trade-off in diffuseness and noisiness. Design 2, on the other hand, achieves a much better SNR (12.38) compared to both Design 1 and the GRIN metalens, which is attributed to its low $J_4$ value. The performance of the optimized designs is benchmarked against the state-of-the-art hybrid metalens by J. Mei et al. [22], which combines GSL and diffraction grating theory. This comparison, detailed in Table 3, is conducted under consistent numerical aperture (NA) and host material. Compared to Design 2, the metalens by J. Mei has a comparable FWHM, efficiency, and intensity enhancement factor per width, but its SNR (8.13) is worse.

For the case of higher NA (NA = 0.994, $F_0 = 1.11\lambda_L$), DWS design 1 achieves an efficiency and intensity enhancement factor of 61.6% and 26.0, respectively, which are 45% higher than those of the GRIN metalens (42.6% and 18.0). Furthermore, design 2 outperforms the GRIN metalens not only in terms of efficiency and intensity enhancement factor but also in SNR. The advantage of the DWS design framework becomes more pronounced at higher NA values, as large phase gradients are required to focus waves at the edges of the metasurface. GRIN metasurfaces are known to exhibit lower efficiency under such high phase gradient conditions. In contrast, the DWS framework holistically optimizes the entire metasurface, accounting for interactions between its components.

### 2. Transmission type elastic Metalens

As the final example, a transmission type metalens is designed using the DWS method, and its performance is verified through both numerical simulation and experimental measurement.

*a. Modelling and Design.* The base material is aluminum (E = 70GPa, $\rho$ = 2700kg/m³, $\nu$ = 0.33), and the frequency is 200kHz. Since the experiment validation is to be performed using a thin plate, the setup is modeled under plane stress condition. The width of the metasurface is $2W = 20.4\lambda_L$, and it is designed to focus a normally incident longitudinal wave at a distance of $F_0 = 1.2\lambda_L$, which corresponds to an NA of 0.9942.

Since the design task is again symmetric, it is sufficient to model half of the domain. The FEM simulation setup is shown in Fig. 17. Apart from the right edge and the bottom edge, the top edge is also surrounded by ALID to simulate an infinite domain.

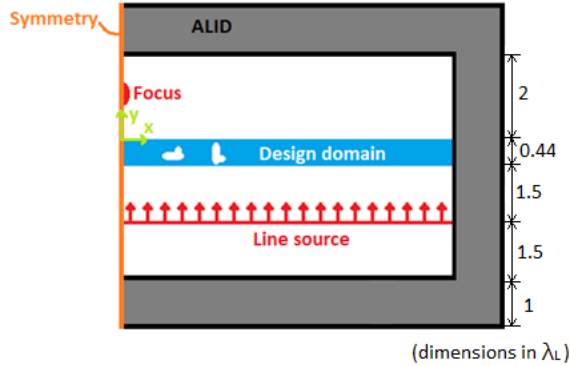

FIG. 17. Schematic of FEM setup used for the design and performance evaluation of a transmitted focusing metasurface.

$$J_1(\boldsymbol{p}) = \text{FWHM}/\lambda_L, \quad (20)$$
$$J_2(\boldsymbol{p}) = |F - F_0|/\lambda_L,$$
$$J_3(\boldsymbol{p}) = -\frac{\int_0^{x_{\min}} |A_L(x, F_0)|^2 dx}{A_I^2 W},$$
$$J_4(\boldsymbol{p}) = -\frac{\int_0^{x_{\min}} |A_L(x, F_0)|^2 dx}{\int_0^{\infty} |A_L(x, F_0)|^2 dx},$$
$$J_5(\boldsymbol{p}) = \frac{\int_D |A_T(x, F_0)|^2 dx}{A_I^2 W},$$
$$J_6(\boldsymbol{p}) = \frac{\int_{D_I} |B_L(x, F_0)|^2 dx}{A_I^2 W},$$

Again, the design domain is subdivided into 24 rectangular subdomains and each subdomain contains 8 ellipses. The objective functions for designing the transmission type metalens are listed in Eq. (20). The first five objectives are the same as those listed in Eq. (15), only that the reflected fields $B_L$ and $B_T$ are replaced by the transmitted wavefield $A_L$ and $A_T$ in $J_3$ through $J_5$. The objective $J_6$ is added to minimize the magnitude of the reflected longitudinal field, where $D_I$ indicates the incident wavefield.

The optimization problem is solved using AGEMOEA, and two designs are selected from the resulting population: (1) the design with the highest efficiency (lowest value for $J_3$), and (2) a design with the lowest objective sum. The geometries of the designs are shown in Fig. 18, and the corresponding parameters are listed in Supplementary Materials.

The designs are compared with a GRIN metalens, the phase profile of the metalens is the same as Eq. (17), but since the wave only propagate through the metalens for only once, the profiles of the material properties are modified as in Eq. (21):

$$\rho(x) = \frac{\lambda_L}{H} \frac{\phi(x)}{2\pi} \rho_0, \quad (21)$$
$$E(x) = \frac{H}{\lambda_L} \frac{2\pi}{\phi(x)} E_0.$$

The performances of our designs and the GRIN metalens are illustrated in Fig. 19. The wavefield slices are shown in Fig. 20, and the quantitative results are listed in Table 5. Again, the DWS designs significantly outperform the GSL-based GRIN metalens, achieving excellent efficiency (53.8% vs. 19.2%) and intensity enhancement factor (37.7 vs. 9.67). This performance is attributed to the DWS designs' effective leverage of nonlocal interactions across the metasurface. Moreover, designs such as Design 2, which offer a better (FWHM and SNR at the cost of reduced efficiency and intensity enhancement, are also available. Such designs are preferable for applications that prioritize FWHM and SNR.

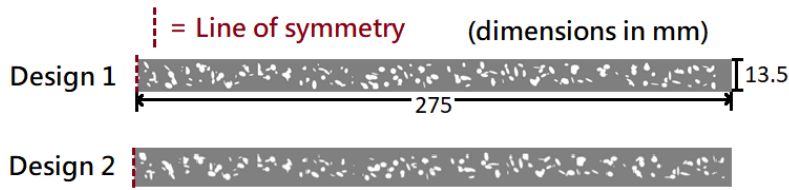

FIG. 18. DWS reflective metalenses designed to focus longitudinal wave at $F_0 = 4\lambda_L$ and $F_0 = 1.1\lambda_L$ respectively.

*Contact author: mewye@ust.hk

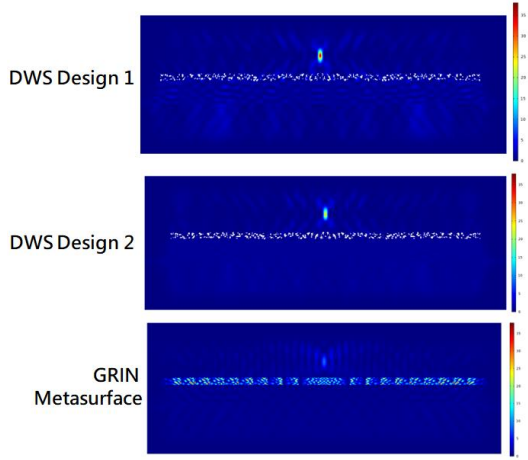

FIG. 19. The normalized intensity fields as a longitudinal wave incident normally from the bottom of the transmission type DWS designs and the comparison with a GRIN metasurface.

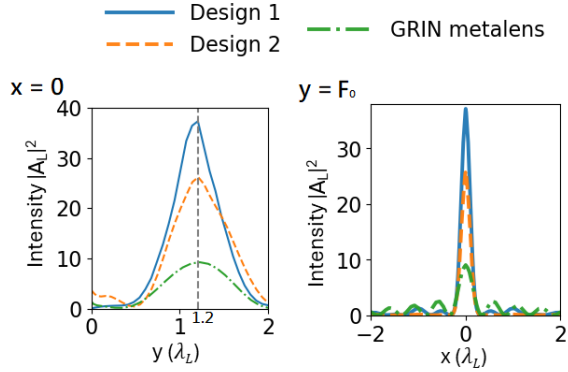

FIG. 20. Slices of the intensity fields along the line of symmetry x = 0 and and focal line y = F0 for the design case (a) $F_0 = 4\lambda_L$ and (b) $F_0 = 1.1\lambda_L$.

TABLE 5. Performance metrics for DWS and GRIN metalenses.

|  | Design 1 | Design 2 | GRIN |
|---|---|---|---|
| FWHM ($\lambda$) | 0.411 | 0.331 | 0.321 |
| Efficiency | 53.8% | 37.9% | 19.2% |
| SNR | 14.9 | 22.7 | 5.63 |
| Intensity enhancement | 37.7 | 26.3 | 9.67 |

*b. Experimental Validation.* Design 1 in Fig. 18 is fabricated on a 1 mm-thick aluminum plate, and performance is experimentally measured using the methodology described in Section II.D. Specifically, the longitudinal-wave amplitude is measured at the following locations relative to the focal point: (0, F0), (0, F0 ± 5 mm), (0, F0 ± 10 mm), (±20 mm, F0), and (±40 mm, F0), where F0 is the y coordinate of the focus point. Thus, five points along the symmetry line at 5 mm spacing around the focus are sampled, and additional points along the focal line at 20 mm spacing around the focus are sampled. Fig. 21 shows a photograph indicating these sensing locations. To quantify focusing performance relative to the incident field, a baseline specimen without the metasurface is also measured to obtain the reference (incident) wave magnitude. This magnitude is then used to normalize the measured wave amplitudes on the specimen with the metasurface.

The results are superimposed on the simulation results in Fig. 22. The experimentally measured intensity enhancement factor is 10.7, which is lower than the simulation prediction of 37.7. Part of this discrepancy can be attributed to differences between the idealized modelling assumptions and the experimental conditions. The simulations are conducted in the frequency domain with no material losses, effectively representing a steady-state harmonic response in which late-arriving reflections from the elliptical hole boundaries can constructively interfere and elevate the peak at the focus. By contrast, the experiment uses a finite-duration toneburst and time-gated analysis that intentionally suppress late arrivals and modal overlap, thereby reducing the measured peak relative to the steady-state prediction. Additional departures arise from the plane-stress assumption. The S0 Lamb mode is treated as a proxy for the plane stress longitudinal wave, whereas dispersion, through-thickness effects, and residual contributions from non-target modes in the actual specimen can diminish the focal amplitude. Finally, material damping, air loading, and minor energy leakage into supports, all present in the laboratory but omitted from the lossless model, further lower the experimental peak.

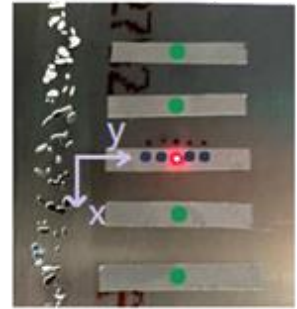

FIG. 21. Positions on the specimen where the amplitude of the longitudinal wave is measured.

*Contact author: mewye@ust.hk

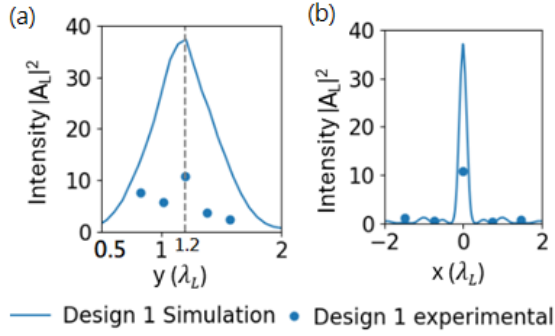

FIG. 22. Simulation predicted and experimentally measured normalized intensity of the longitudinal wavefield along the (a) line of symmetry and (b) the focal line.

Nevertheless, the measurements demonstrate clear focusing at the intended location, and the experimentally obtained enhancement factor is higher than the predicted enhancement of a GRIN metalens designed with the ideal material profile (9.67). This confirms that Design 1 achieves the desired wave concentration in practice.

## V. CONCLUSIONS

In this work, we have proposed a DWS framework for the design of high-efficient elastic metasurfaces capable of performing arbitrary, complex wave control functions. This general-purpose approach significantly reduces the number of design variables and facilitates the creation of monolithic structures that exploit non-local interactions—a key mechanism for achieving high performance. By formulating design objectives directly based on wavefields, the framework enables the relatively straightforward realization of high-performance metasurfaces. Furthermore, the integration of multiple physical objectives into the optimization process not only enhances device performance but also offers deeper insight into the inherent trade-offs in metasurface design.

The framework's effectiveness was demonstrated through several examples of simultaneous mode conversion with beam steering and wave focusing, achieving significant improvements in efficiency and performance compared to conventional methods. Specifically, for single-frequency beam steering, it achieved high efficiencies of over 91%, significantly outperforming GSL-based designs, which suffered from low efficiency and wave scattering due to a neglect of non-local interactions. The framework's holistic optimization approach also enabled the creation of a frequency-multiplexed metasurface that successfully operated at two distinct frequencies, efficiently steering waves into different angles. Furthermore, when designing focusing metasurfaces, the framework produced reflective and transmission type lenses with high efficiency, even at very high numerical apertures where traditional GRIN designs fail. It outperformed both GRIN lenses and a state-of-the-art hybrid design, achieving markedly higher focusing efficiencies and demonstrating its ability to overcome fundamental limitations of existing methodologies.

While the metasurfaces designed using the DWS framework demonstrate high power efficiency, the resulting geometries include cusps and sharp corners, which may complicate fabrication or lead to stress concentration issues. A promising approach to address this issue is to integrate a Gaussian filter into the optimization process. This approach produces smoother structural geometries while preserving or even improving device performance, thereby facilitating the development of more efficient and manufacturable metasurfaces.

## ACKNOWLEDGMENTS

We thank Prof. Fan Shi for providing access to the Polytec OFV-505 laser Doppler vibrometer for our experimental measurements.

*Contact author: mewye@ust.hk

*Contact author: mewye@ust.hk